\documentclass[twocolumn,showpacs,amssymb,aps,prl,floatfix]{revtex4}

\usepackage{epsfig}
\usepackage{amsmath}

\newcommand{\be}{\begin{equation}}
\newcommand{\ee}{\end{equation}}
\newcommand{\bea}{\begin{eqnarray}}
\newcommand{\eea}{\end{eqnarray}}
\newcommand{\bean}{\begin{eqnarray*}}
\newcommand{\eean}{\end{eqnarray*}}

\newcommand{\om} {\omega}

\newcommand{\pv}{{\mathbf p}}

\begin{document}

\title{Bottomonium above deconfinement in lattice nonrelativistic QCD}

\author{G.~Aarts${}^a$, S.~Kim${}^b$,  M.~P.~Lombardo${}^c$, 
M.~B.~Oktay${}^d$, S.~M.~Ryan${}^e$, D.~K.~Sinclair${}^f$ 
and  J.-I.~Skullerud${}^g$}

\affiliation{
${}^a$Department of Physics, Swansea University, Swansea SA2 8PP, United Kingdom \\
${}^b$Department of Physics, Sejong University, Seoul 143-747, Korea \\
${}^c$INFN-Laboratori Nazionali de Frascati, I-00044, Frascati (RM) Italy \\
${}^d$Physics Department, University of Utah, Salt Lake City,  Utah, USA \\
${}^e$School of Mathematics, Trinity College, Dublin 2, Ireland \\
${}^f$HEP Division, Argonne National Laboratory, 9700 South Cass Avenue,
 Argonne, Illinois 60439, USA\\
${}^g$Department of Mathematical Physics, National University of
  Ireland Maynooth, Maynooth, County Kildare, Ireland
}

\date{\today}

\begin{abstract}

We study the temperature dependence of bottomonium for temperatures in the 
range $0.4 T_c < T < 2.1 T_c$, using nonrelativistic dynamics for the bottom 
quark and full relativistic lattice QCD simulations for $N_f=2$ light 
flavors on a highly anisotropic lattice. We find that the $\Upsilon$ is 
insensitive to the temperature in this range, while the $\chi_b$ 
propagators show a crossover from the exponential decay characterizing the 
hadronic phase to a power-law behaviour consistent with nearly-free 
dynamics at $T \simeq 2 T_c$.

\end{abstract}


\maketitle


{\em Introduction --} 
 Heavy quark bound states are important probes of the dynamics in the 
Quark Gluon Plasma: charmonium suppression 
\cite{Matsui:1986dk,Satz:2006kba} has been observed at a variety of 
energies at SPS \cite{Arnaldi:2009ph} and RHIC \cite{Adare:2008sh}. While 
the melting of bound states certainly reduces quarkonium production, the 
converse is not necessarily true:  different, even competing, effects make 
it difficult to interpret charmonium suppression patterns.  It has been 
noted that such effects should be less significant for bottomonium (see 
e.g.\ Ref.\ \cite{Rapp:2008tf} for a review). 
Since at LHC energies bottomonium will be produced 
copiously \cite{Lansberg:2008zm,Alessandro:2006yt}, 
precision studies of the suppression pattern 
and its unambiguous link with the spectrum of bound states should be 
possible. The advent of the LHC calls therefore for precision studies of 
bottomonium at high temperature.

 Due to the large mass of the bottom quark, it is customary to study the 
bottomonium spectrum at zero temperature using nonrelativistic QCD (NRQCD) 
\cite{Lepage:1992tx,Davies:1994mp,Bodwin:1994jh} and other effective field 
theories \cite{Brambilla:2004jw}. In this Letter, we employ NRQCD to study 
the response of bottomonium to a thermal medium of quarks and gluons in 
the temperature range $0.4T_c < T < 2.1T_c$, at the onset of the initial 
temperature attained in heavy ion collisions at the LHC 
\cite{Wiedemann:2009sa}. We use dynamical anisotropic lattice 
configurations with two light flavors, which have been exploited before 
in a relativistic study of charmonium \cite{Aarts:2007pk,Morrin:2006tf}. 
As explained below, the use of NRQCD is a controlled approach which avoids 
many of the unwanted systematic effects encountered when using 
relativistic dynamics for bottomonium at nonzero temperature 
\cite{Jakovac:2006sf,Petreczky:2010yn,Rapp:2009my}.
 An early study of bottomonium at nonzero temperature using NRQCD (on a
quenched background for the ${}^1S_0$ and ${}^3S_1$ states only) can
be found in the pioneering work \cite{Fingberg:1997qd}.

{\em NRQCD at nonzero temperature -- }
 In contrast to the case at zero temperature \cite{Brambilla:2004jw}, the 
use of potential models to analyse quarkonium at nonzero temperature is 
less well defined due to the uncertainty about which potential to use (see 
e.g.\ Ref.\ \cite{Mocsy:2007yj} and references therein). Recently this has 
been clarified by casting the problem in the language of effective field 
theory at nonzero temperature 
\cite{Laine:2006ns,Laine:2007gj,Burnier:2007qm,Beraudo:2007ky,Brambilla:2008cx,Brambilla:2010vq}. 
The series of effective field theories that is obtained is based on the 
hierarchy $M\gg T> g^2M> gT \gg g^4M$, where $M$ is the heavy quark mass 
and $g$ is the gauge coupling. Ref. \cite{Laine:2008cf} provides a clear 
introduction. Integrating out thermal degrees of freedom generates an 
imaginary part for the interquark potential, which highlights the absence 
of stable states once they are immersed in a thermal medium. Limitations 
of approaches based on potential models and the Schr\"odinger equation are 
discussed in Ref.\ \cite{Beraudo:2010tw}.

In the effective thermal field theory setup 
\cite{Laine:2006ns,Laine:2007gj,Burnier:2007qm,Beraudo:2007ky,Brambilla:2008cx,Brambilla:2010vq,Laine:2008cf}, NRQCD is 
the first theory obtained when integrating out ultraviolet degrees of freedom. 
We study this theory nonperturbatively on the lattice and therefore do not 
require weak-coupling arguments as in the hierarchy of effective field 
theories alluded to above. Since NRQCD relies on the scale separation 
$M\gg T$ and we study temperatures up to $2 T_c \simeq 400$ MeV, its 
application is fully justified.

NRQCD has an additional advantage. At nonzero temperature, spectroscopy for 
relativistic quarks is hindered by the periodicity of the lattice in the 
temporal direction and the reflection symmetry of mesonic correlators, 
visible in, e.g., the standard relation between a correlation function and 
its spectral function,
 \be
 \label{eqK}
 G(\tau) = \int_0^\infty \frac{d\omega}{\pi}\, 
 \frac{\cosh\left[\omega(\tau-1/2T)\right]}{\sinh\left(\omega/2T\right)} 
 \rho(\omega).
 \ee
 Nontrivial spectral weight at small $\omega$ yields a constant 
$\tau$-independent contribution to the correlator, which must 
be treated with care \cite{Aarts:2002cc,Aarts:2007wj}. Moreover, it has 
been shown that this contribution can interfere with meson spectroscopy 
\cite{Umeda:2007hy}, which has cast doubt on the status of results for the 
melting or survival of charmonium at high temperature \cite{Petreczky:2008px,
Umeda:2007hy}. 

In NRQCD these problems are not present. Writing $\om=2M+\om'$ and dropping terms that 
are exponentially suppressed when $M\gg T$ \cite{Burnier:2007qm}, the 
spectral relation (\ref{eqK}) reduces to
 \be
 \label{eq:Gnr}
 G(\tau) =
 \int_{-2M}^\infty\frac{d\omega'}{\pi}\, \exp(-\om'\tau) \rho(\omega')
 \;\;\;\;\;\;(\mbox{NRQCD}), 
 \ee
 even at nonzero temperature. As a result, all problems associated with
thermal boundary conditions are absent.

To study what to expect when quarks are no longer bound, consider 
free quarks in continuum NRQCD with energy $E_\pv=\pv^2/2M$. The 
correlators for the $S$ and $P$ waves are then of the form 
\cite{Burnier:2007qm}
 \begin{align}
\label{eq:GS}
 G_{S}(\tau)  \sim& \int \frac{d^3p}{(2\pi)^3}\, \exp(-2E_\pv\tau) 
  \sim \tau^{-3/2}, \\ 
\label{eq:GP}
 G_{P}(\tau)  \sim & \int \frac{d^3p}{(2\pi)^3}\, \pv^2 \exp(-2E_\pv\tau)  
  \sim \tau^{-5/2},
\end{align}
i.e., they decay as a power for large euclidean time. Of course,
interactions and finite lattice spacing and volume effects are expected
to modify this in the realistic case.

\begin{table}[b]
\begin{tabular}{llcccc}
\hline\hline \multicolumn{1}{c}{$N_s$} & \multicolumn{1}{c}{$N_\tau$} &
\multicolumn{1}{c}{$a^{-1}_\tau$} & \multicolumn{1}{c}{$T$(MeV)} &
\multicolumn{1}{c}{$T/T_c$} & \multicolumn{1}{c}{No.\ of Conf.}
\\ \hline 12 & 80 & 7.35GeV &  90 & 0.42 &  74 \\      12 & 32 &
7.06GeV & 221 & 1.05 & 500 \\ 12 & 24 & 7.06GeV & 294 & 1.40 & 500
\\ 12 & 16 & 7.06GeV & 441 & 2.09 & 500 \\ \hline
\end{tabular}
\label{tab:latticedetail}
 \caption{Summary of the lattice data set. The lattice spacing is set 
using the $1P-1S$ spin-averaged splitting in 
charmonium~\cite{Oktay:2010tf}.}  
\end{table}

{\em Lattice simulations --} Gauge configurations 
with two degenerate dynamical light Wilson-type quark flavors are produced on 
highly anisotropic lattices ($\xi\equiv a_s/a_\tau=6$) of size 
$N_s^3\times N_\tau$. 
A summary of the lattice datasets is given in Table~I, while more 
details of the lattice action and parameters can be found in 
Refs.~\cite{Aarts:2007pk,Morrin:2006tf}. We computed NRQCD propagators on these 
configurations using a mean-field improved action with tree-level coefficients, which includes
terms up to and including ${\cal O}(v^4)$,
where $v$ is the typical velocity of a bottom quark in bottomonium
(see Ref.\ \cite{Davies:1998im} for a discussion of the systematics). 
The states we consider are listed in Table II.

An accurate determination of bottomonium spectroscopy requires careful 
tuning of the bare heavy quark mass $m_b$ to satisfy NRQCD dispersion 
relations \cite{Davies:1994mp}. Since the main goal of this work is to 
study the finite-temperature modification of NRQCD propagators, an approximate 
choice of $a_sm_b$ is made such that $m_b \simeq 5$ GeV.

\begin{table}[t]
\begin{tabular}{lccc}
 \hline\hline 
 \multicolumn{1}{l}{state} & \multicolumn{1}{c}{$a_\tau\Delta E$} 
 & \multicolumn{1}{c}{Mass (MeV)} & \multicolumn{1}{c}{Exp.\ (MeV) \cite{Nakamura:2010zz} } \\ 
 \hline 
 1$^1S_0 (\eta_b)    $ & 0.118(1) &  9438(7)  &  9390.9(2.8)  \\ 
 2$^1S_0 (\eta_b(2S))$ & 0.197(2) & 10009(14) &  -            \\ 
 1$^3S_1 (\Upsilon)  $ & 0.121(1) &  9460$^*$ &  9460.30(26)  \\ 
 2$^3S_1 (\Upsilon') $ & 0.198(2) & 10017(14) & 10023.26(31)  \\ 
 1$^1P_1 (h_b)       $ & 0.178(2) &  9872(14) & -              \\ 
 1$^3P_0 (\chi_{b0}) $ & 0.175(4) &  9850(28) &  9859.44(42)(31) \\ 
 1$^3P_1 (\chi_{b1}) $ & 0.176(3) &  9858(21) &  9892.78(26)(31) \\ 
 1$^3P_2 (\chi_{b2}) $ & 0.182(3) &  9901(21) &  9912.21(26)(31)  \\  
 \hline
 \end{tabular}
 \caption{Zero temperature bottomonium spectroscopy from NRQCD. The 
$1^3S_1 (\Upsilon)$ state is used to set the scale.
}
\label{tab:bindE}
\end{table}

{\em Zero temperature results --} The zero temperature spectrum from our 
analysis is summarized in Table~\ref{tab:bindE}.  This spectrum is 
obtained using a combination of point and extended sources in the 
different channels, in order to extract both the ground state and the 
first excited state. 

Because level splittings are relatively insensitive to $m_b$ and to 
avoid the difficulties in calculating the rest mass in NRQCD, we have 
combined $m_\Upsilon(1^3S_1)=9460$ MeV 
(from the Particle Data Book \cite{Nakamura:2010zz}) with the mass 
splittings obtained from our results to predict the bottomonium spectrum.

As can be seen from Table~\ref{tab:bindE}, the zero 
temperature spectrum is reproduced reasonably well.  The hyperfine 
splitting between the $\eta_b ({}^1S_0)$ and $\Upsilon ({}^3S_1)$ is much
smaller than the experimental value, due to the coarse spatial 
lattice spacing, the use of tree-level coefficients, the relatively
heavy sea quarks and contributions of higher order in $v^2$
(see Ref.\ \cite{Davies:1998im} for earlier calculations and discussions).
In this study we are primarily interested in qualitative changes to the 
$\Upsilon$ and $\chi_b$ correlators as the temperature changes, 
so a precision determination is not a major concern.

\begin{figure}[t]
 \centerline{\includegraphics*[width=8.5cm]{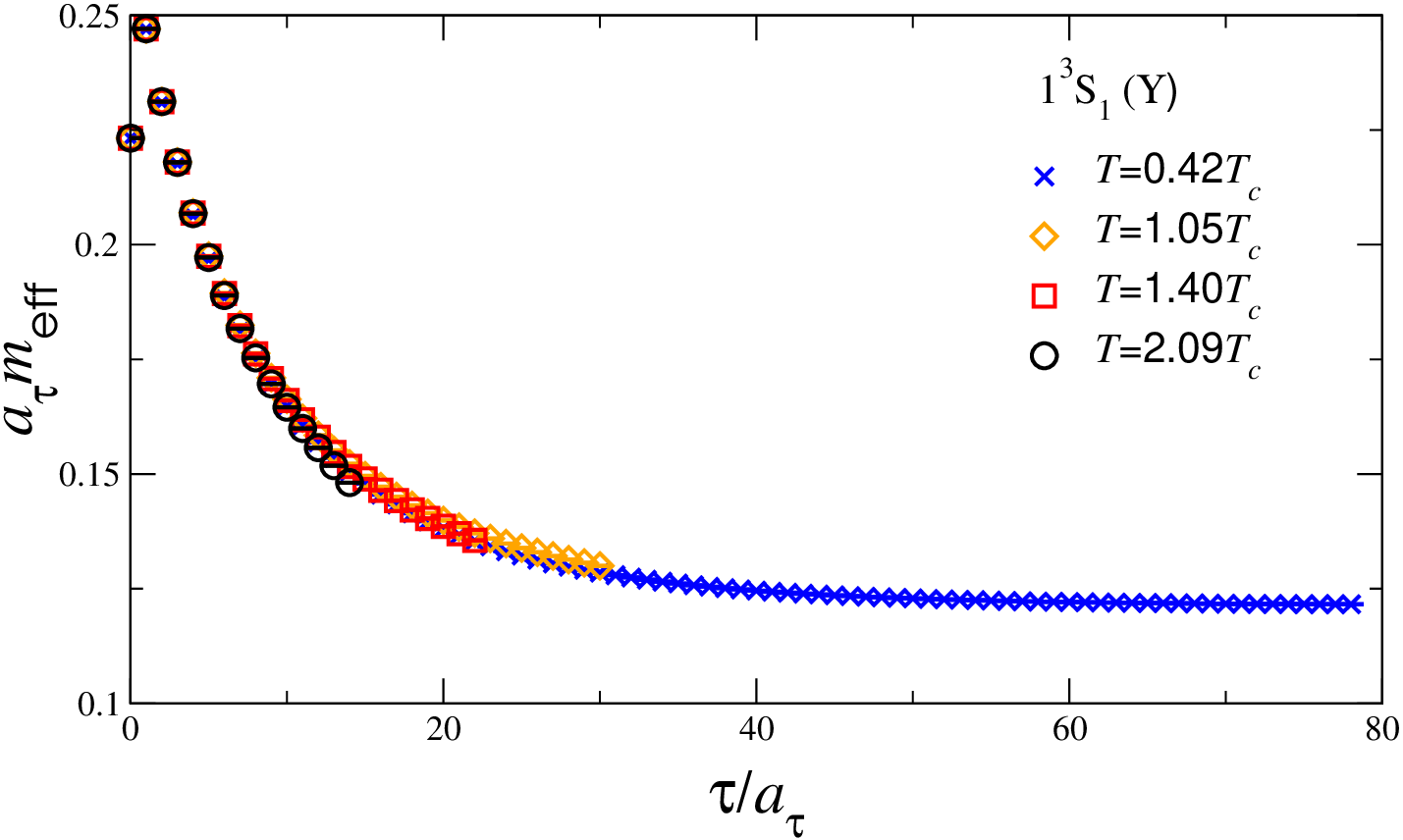}}
 \centerline{\includegraphics*[width=8.5cm]{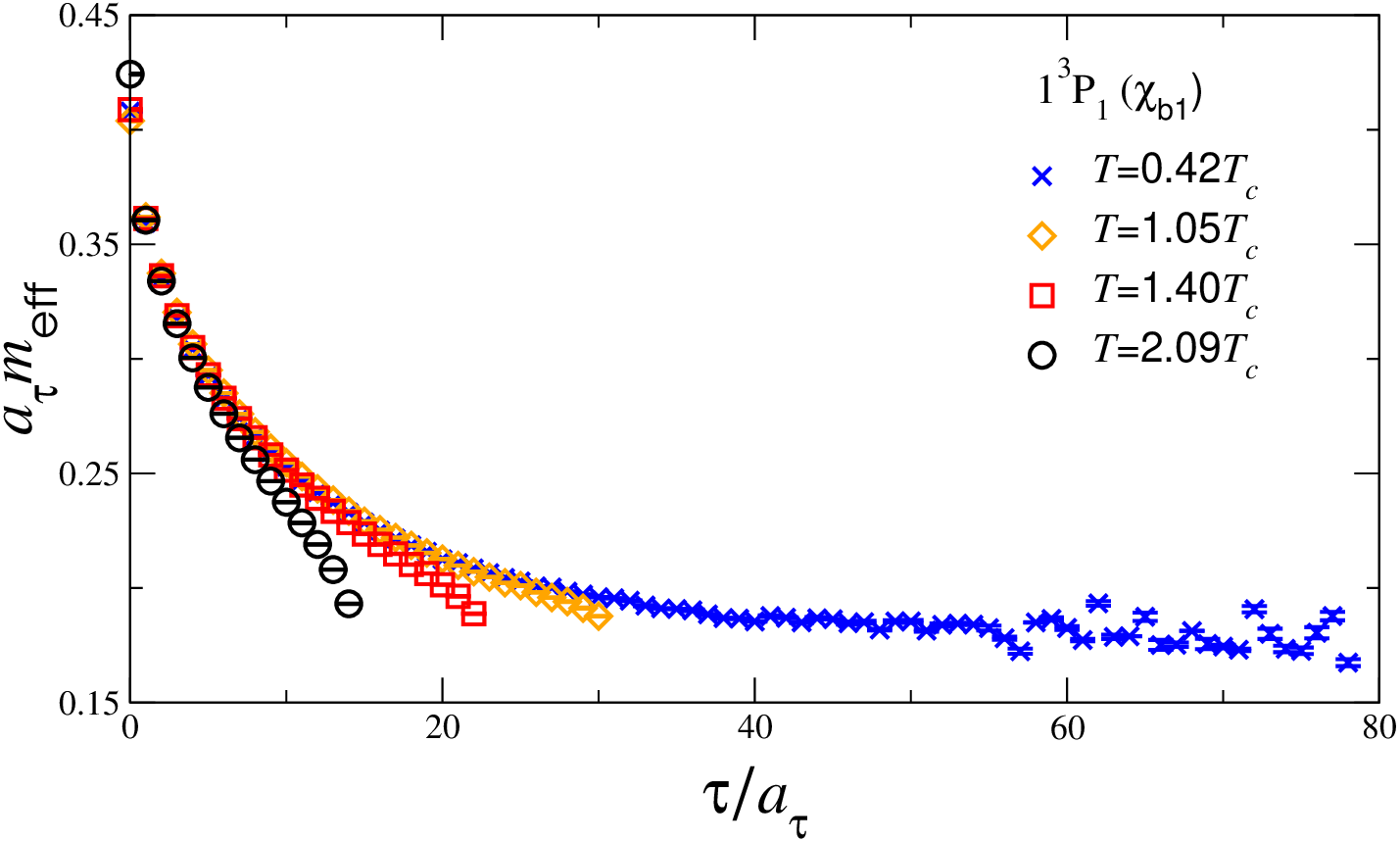}}
 \caption{
 Effective mass plots for the $\Upsilon$ (above) and  $\chi_{b1}$ 
(below) using point sources for various temperatures: note the different 
temperature dependence.} 
 \label{fig:effmass}
 \end{figure}

\begin{figure}[t]
  \centerline{\includegraphics*[width=8.5cm]{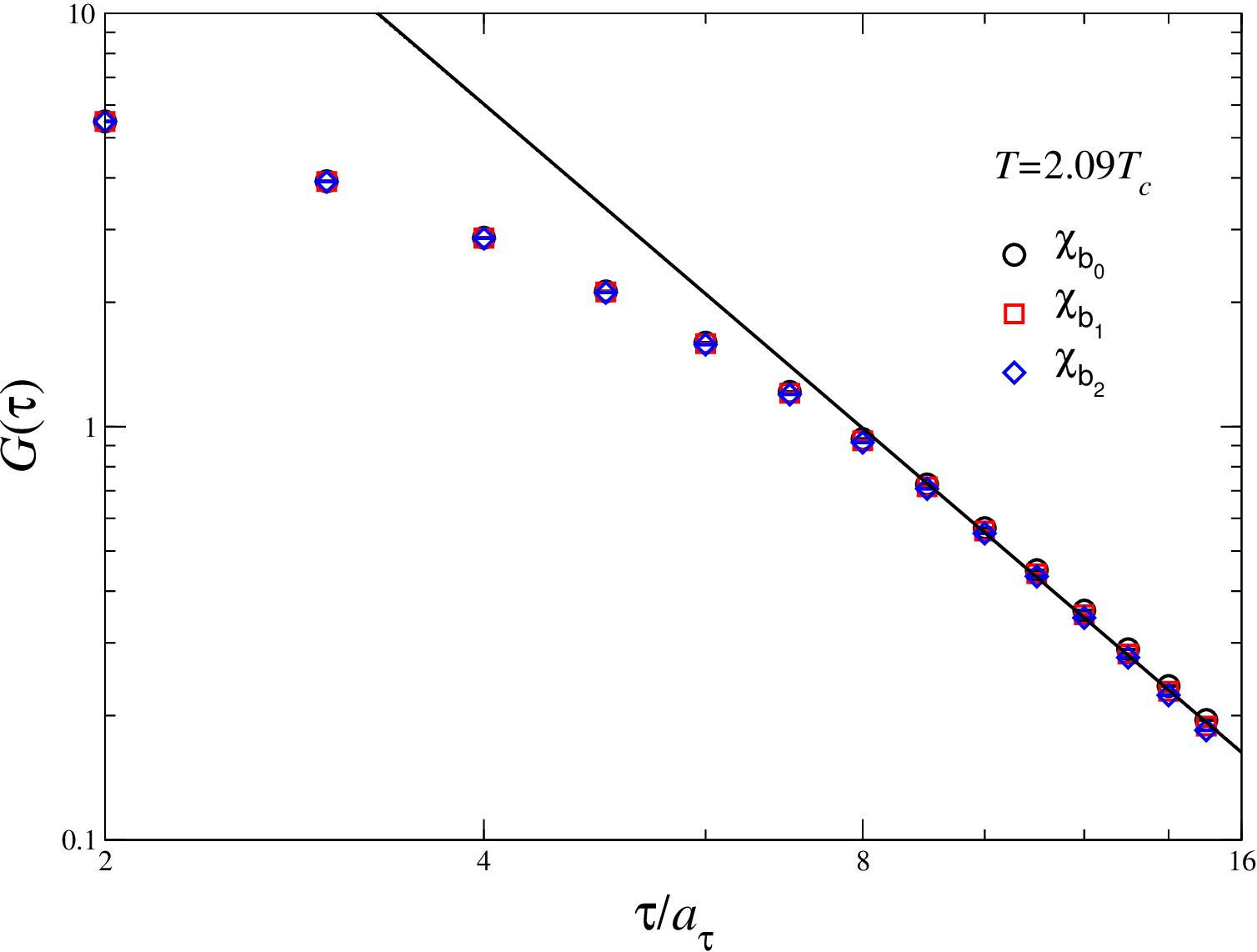}}
  \caption{
  $\chi_{b}$ propagators on a log-log scale, at the highest temperature 
$T=2.09T_c$. The straight line is a fit to $G(\tau) = c\tau^{-d}$, with 
$c=223.2\pm 0.5$ and $d=2.605\pm 0.001$, using $\tau/a_\tau=10,\ldots,15$.
 } 
 \label{fig:Fig2_log_log} 
 \end{figure}

 \begin{figure}[t]
 \centerline{\includegraphics[width=8cm]{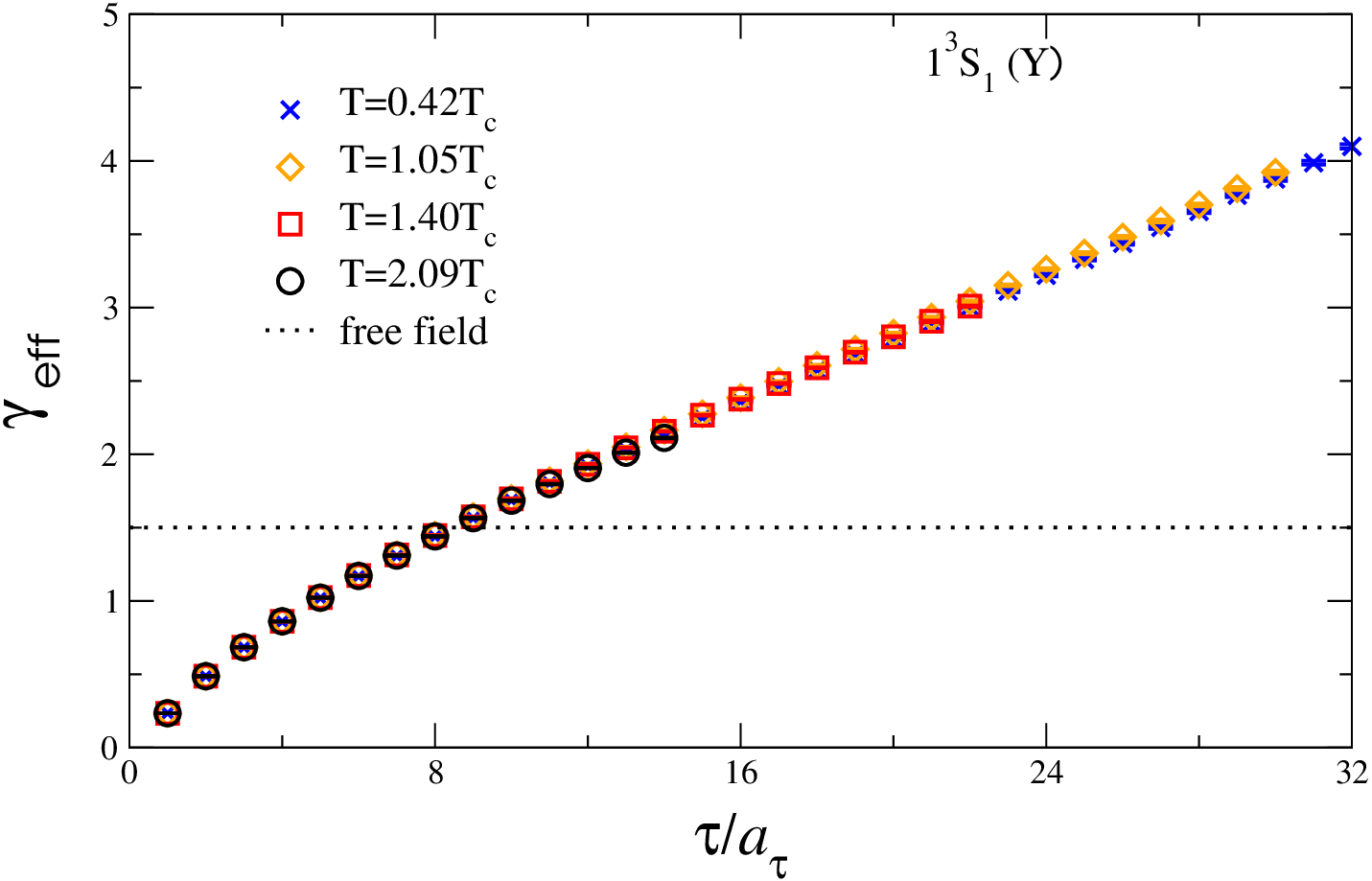}}
 \centerline{\includegraphics[width=8cm]{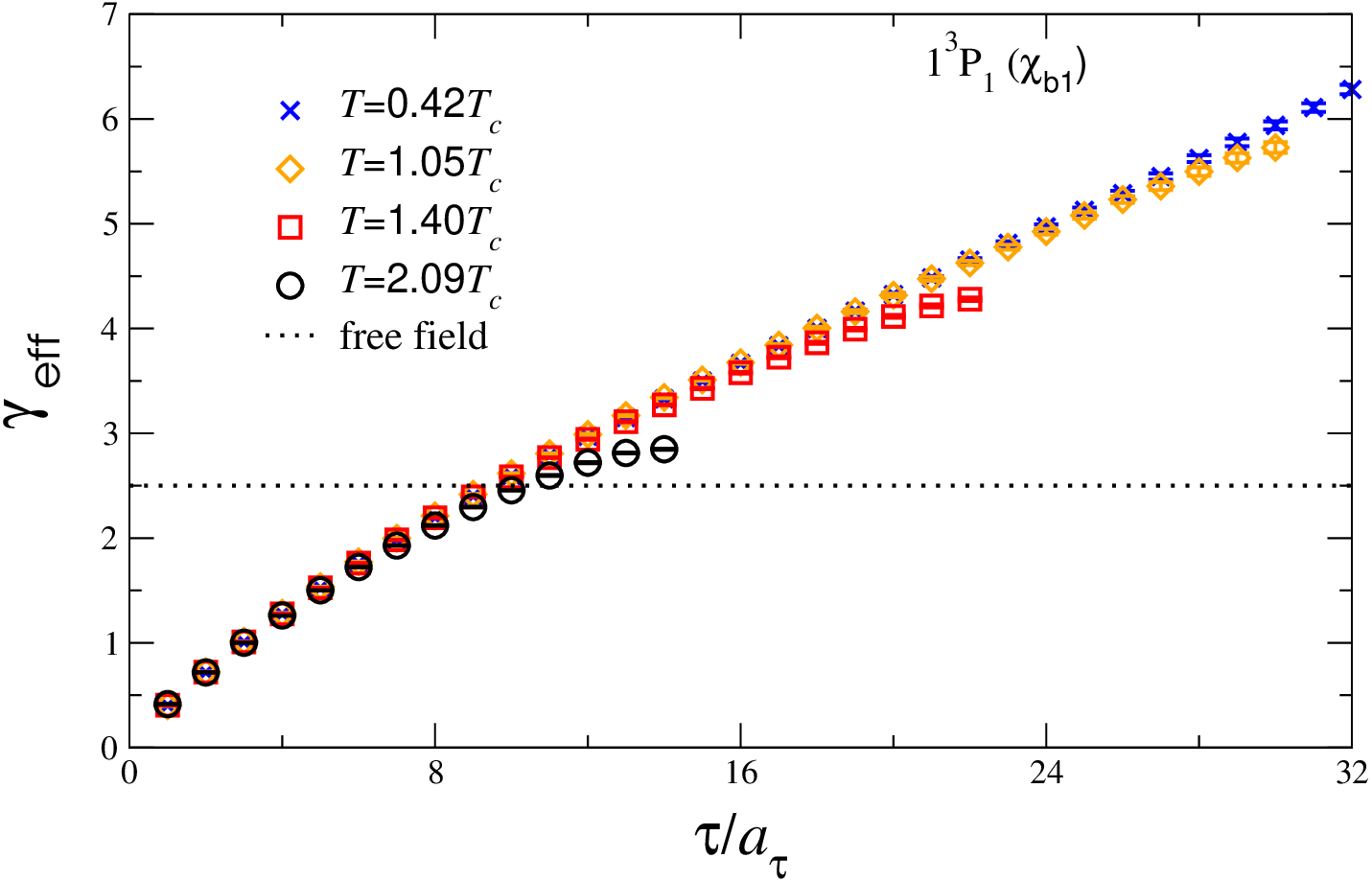}}
 \caption{
 Effective exponents $\gamma_{\rm eff}(\tau)$ for the $\Upsilon$ (above)
and $\chi_{b1}$ (below), as a function of Euclidean time for various
temperatures. The dotted line indicates the noninteracting 
result in the continuum.
 } 
 \label{fig:power}
 \end{figure}

{\em $\Upsilon$ and $\chi_b$ in the plasma --} To investigate thermal 
effects, we focus on the $\Upsilon$ and $\chi_b$ states, computed with 
point sources. Following the discussion above, our aim is to see a 
transition from exponential decay in the hadronic phase, $G(\tau)\sim 
\exp(-\Delta E\tau)$, characterizing bound states, to 
power law decay, $G(\tau) \sim \tau^{-\gamma}$, see Eqs.\ 
(\ref{eq:GS}, \ref{eq:GP}), characterizing quasi-free behaviour.

In Fig.\ 1, standard effective masses, defined by 
 \begin{equation}
 m_{\rm eff}(\tau) = -\log[G(\tau)/G(\tau-a_\tau)],
 \end{equation} 
 are shown for both the $\Upsilon$ and the $\chi_{b1}$ propagators at 
various temperatures. Single exponential decay should yield a 
$\tau$-independent plateau. In both cases we find that at the lowest 
temperature, $T=0.42T_c$, exponential behaviour is visible provided one 
goes to late euclidean times.
 Relevant for the topic of this Letter is that in the case of the 
$\Upsilon$, the data at the higher temperatures do not show any 
significant deviation from the low-temperature result. On the other hand 
for the $\chi_{b1}$ a strong temperature dependence is visible, especially 
at the two highest temperatures, ruling out pure exponential long time 
decay.
 We take this as a first indication that the $\Upsilon$ is not sensitive 
to the quark-gluon plasma up to $T\simeq 2T_c$, while the $\chi_b$ may 
melt at much lower temperatures. 

To investigate the behavior of the $\chi_b$ propagators in more detail, we 
display in Fig.\ 2 the $\chi_{b0}$, $\chi_{b1}$ and $\chi_{b2}$ 
propagators on a log-log scale, at the highest temperature. The straight 
line is a fit of the form $G(\tau) = c \tau^{-d}$, which is motivated by 
the continuum expression in the absence of interactions (\ref{eq:GP}). We 
conclude that a power decay describes the data well at large euclidean time, 
with a power $d=2.605(1)$ which is close to the continuum noninteracting 
value of $5/2$.

To visualize the approach to quasi-free behaviour in another way, 
we construct effective power plots, using the definition 
 \begin{equation}
 \label{eq:power}
 \gamma_{\rm eff}(\tau) = -\tau\frac{G'(\tau)}{G(\tau)}
 = -\tau\frac{G(\tau+a_\tau) - G(\tau -a_\tau)}{2a_\tau G(\tau)},
 \end{equation}
 where the prime denotes the (discretized) derivative.  For a power decay, 
$G(\tau)\sim \tau^{-\gamma}$, this yields a constant result, $\gamma_{\rm 
eff}(\tau)=\gamma$. On the other hand, for an exponential decay, 
$G(\tau)\sim \exp(-\Delta E\tau)$, this yields a linearly rising result, 
$\gamma_{\rm eff}(\tau)=\Delta E\tau$. The results are shown in Fig.\ 3. 
We confirm again that the $\Upsilon$ displays essentially no temperature 
dependence, while for the $\chi_{b1}$ we observe a tendency to flatten 
out, corresponding to power decay at large euclidean time. 
Also shown are the effective exponents in the continuum noninteracting limit. 
In the case of the $\chi_{b1}$, we observe that the 
effective exponent tends towards the noninteracting result at the highest 
temperature we consider.

{\em Summary --} 
We have studied the behavior of $\Upsilon$ and $\chi_b$ at high
temperature using anisotropic lattice simulations. The bottom quark is 
treated via NRQCD and the light quark dynamics is realized by exploiting
lattice configurations with two flavors of dynamical quarks.  The
nonrelativistic approximation for the bottom quark is well justified
in the range of temperatures we have explored, and has many technical
advantages.
 At nonzero temperature, the main benefit of using NRQCD over the standard 
relativistic formulation is that the only temperature dependence in the 
NRQCD correlators is due to the thermal medium, and not due to thermal 
boundary conditions. We found that this offers a 
much cleaner signal for the crossover between bound and melted states. It 
also improves the prospects for extracting spectral functions inverting 
Eq.\ (\ref{eq:Gnr}) using the Maximal Entropy Method.  This is 
currently in progress. 
It will also be interesting to compare the results presented here 
with those obtained using a relativistic treatment of bottom quarks, 
employing the same anisotropic action as was used for charmonium 
\cite{Aarts:2007pk,Oktay:2010tf}.  This will give an estimate of the
possible systematic uncertainties inherent in the two formulations.

Our results indicate that the $\Upsilon$ shows no temperature dependence 
up to $2.09T_c$, while the $\chi_b$ propagators are sensitive to the 
presence of the thermal medium immediately above $T_c$. Power-law 
decay of the $\chi_b$ propagators is visible at $T=1.4T_c$, while at the 
highest temperature studied, $T \simeq 2T_c$, we found consistency with 
nearly-free dynamics.
 The effective power, defined in Eq.\ (\ref{eq:power}), is temperature 
dependent and approaches the noninteracting result at the highest 
temperature we considered. It would be interesting to understand the
temperature dependence and crossover between exponential and power 
decay analytically, within the framework of effective field theories 
mentioned above.

\vspace{0.2 truecm}
\begin{acknowledgments}
MPL thanks Helmut Satz, Peter Petreczky and participants in the ``In Media'' 
group of the Quarkonium Working Group  for 
fruitful discussions. SK and MPL thank the Yukawa Institute of Theoretical 
Physics, Kyoto, and   GA and  MPL thank Trinity College Dublin and the National University of  Ireland Maynooth, for their hospitality.
SMR and JIS are grateful to the Trinity Centre for High-Performance Computing
for their support.
 GA is supported by STFC. 
 SK is supported by the National Research Foundation of Korea grant funded 
by the Korea government (MEST) No.\ 2010-0022219.
 DKS is supported in part by US Department of Energy contract 
DE-AC02-06CH11357.
 JIS is supported by Science Foundation Ireland grant 08-RFP-PHY1462.
 SMR is supported by the Research Executive Agency (REA) of the European Union 
under Grant Agreement number PITN-GA-2009-238353 (ITN STRONGnet). 
\end{acknowledgments}

\end{document}